\documentclass[twocolumn]{aastex63}
\usepackage{amsmath}
\usepackage{graphics}
\usepackage{epstopdf}

\usepackage{hyperref}

\newcommand{\del}{\partial}

\newcommand{\addsf}[1]{{\color{magenta} #1}} % added by SF

\newcommand{\addkt}[1]{{\color{red} #1}} % added by KT
\newcommand{\addms}[1]{{\color{blue} #1}} % added by MS

\renewcommand{\addsf}[1]{{\color{black} #1}} % added by SF
\renewcommand{\addkt}[1]{{\color{black} #1}} % added by KT
\renewcommand{\addms}[1]{{\color{black} #1}} % added by MS

\received{}
\revised{\today}
\accepted{}

\shorttitle{Ultra-delayed explosion}
\shortauthors{Fujibayashi et al.}

\begin{document}

\title{Ultra-delayed neutrino-driven explosion of rotating massive-star collapse}

\correspondingauthor{Sho Fujibayashi}
\email{sho.fujibayashi@aei.mpg.de}

\author[0000-0001-6467-4969]{Sho Fujibayashi}
\affiliation{Max-Planck-Institut f\"ur Gravitationsphysik (Albert-Einstein-Institut), Am M\"uhlenberg 1, D-14476 Potsdam-Golm, Germany}

\author[0000-0002-6705-6303]{Koh Takahashi}
\affiliation{Max-Planck-Institut f\"ur Gravitationsphysik (Albert-Einstein-Institut), Am M\"uhlenberg 1, D-14476 Potsdam-Golm, Germany}

\author[0000-0002-2648-3835]{Yuichiro Sekiguchi}
\affiliation{Center for Gravitational Physics, Yukawa Institute for Theoretical Physics, Kyoto University, Kyoto, 606-8502, Japan}
\affiliation{Department of Physics, Toho University, Funabashi, Chiba 274-8510, Japan}

\author[0000-0002-4979-5671]{Masaru Shibata}
\affiliation{Max-Planck-Institut f\"ur Gravitationsphysik (Albert-Einstein-Institut), Am M\"uhlenberg 1, D-14476 Potsdam-Golm, Germany}
\affiliation{Center for Gravitational Physics, Yukawa Institute for Theoretical Physics, Kyoto University, Kyoto, 606-8502, Japan}

\begin{abstract}
  Long-term neutrino-radiation hydrodynamics simulations in full
  general relativity are performed for the collapse of 
  rotating massive stars that are evolved from He-stars with their initial
  mass of $20$ and $32M_\odot$. It is shown that if the collapsing stellar core has
  sufficient angular momentum, the rotationally-supported
  proto-neutron star (PNS) survives for seconds accompanying the formation
  of a massive torus of mass larger than $1\,M_\odot$.  Subsequent mass
  accretion onto the central region produces a massive and compact
  central object, and eventually enhances the neutrino luminosity beyond
  $10^{53}$\,erg/s, resulting in a very delayed neutrino-driven explosion
  in particular toward the polar direction.  The kinetic energy of the
  explosion can be appreciably higher than $10^{52}$\,erg for a massive progenitor star and compatible with that of energetic supernovae like broad-line type-Ic supernovae. 
  By the subsequent accretion, the massive PNS collapses eventually into a rapidly spinning
  black hole, which could be a central engine for gamma-ray bursts if a massive torus surrounds it.
\end{abstract} 


\section{Introduction}
\label{sec:intro}
Core-collapse supernovae (SNe) are explosive events that
occur at the final stage of the massive-star evolution. In the typical
scenario \citep[e.g.,][]{Janka2012b}, after the collapse of
the iron core of progenitor stars, a proto-neutron star (PNS) is
first formed.  Then, a shock wave is generated at the inner
core of the PNS and propagates outward sweeping the matter. However, because of the photo-dissociation of irons,
the shock is stalled in the middle of the propagation. Subsequently,
the heating by neutrinos emitted from the PNS is believed to play a key role
for supplying the energy to the stalled shock~\citep{Bethe1985a}. If
the neutrino heating timescale becomes shorter than that of the matter
infall from the outer envelop, the stalled shock is revived and 
the explosion is driven by the neutrino
heating~\citep{Janka2001a}.  By contrast, if the neutrino
heating is not efficient enough, the stalled shock eventually goes
back to the PNS and a black hole (BH) is formed.  In particular, for
high-mass progenitor stars with the zero-age main-sequence (ZAMS) mass
of $M_{\rm ZAMS} \agt 40M_\odot$~\citep{Woosley2002a}, the naive expectation for the final fate is the formation of a BH without the shock revival.

As summarized above, the key quantity for the successful 
explosion is the efficiency of the neutrino heating~\citep{Janka2001a}.
In fact, many sophisticated simulations for core-collapse SNe have shown that the
success of the SN explosion depends sensitively on the neutrino
luminosity and neutrino heating efficiency~\citep[for the latest numerical simulations in this field, see, e.g.,][]{Mueller2012,Burrows2019,Nakamura2019,Mezzacappa2020,Muller2020a,Stockinger2020,Kuroda2020,Harada2020,Obergaulinger2020,Bollig2020}.

In this paper, we propose a mechanism by which the neutrino luminosity of the central object is naturally enhanced for very high-mass rotating progenitor stars.
We consider a rotating progenitor core, which results in a PNS rapidly rotating with the rotational period of $\leq 1$\,ms and surrounded by a massive torus with the mass beyond $1M_\odot$. Due to the rapid rotation, the PNS with the rest mass $\agt 3M_\odot$ can survive for the equation of state (EOS) with which the maximum gravitational mass for
cold non-rotating neutron stars (NSs), $M_\mathrm{max}$, is larger than
$2M_\odot$. This appreciably increases the lifetime of the PNS. In
addition, due to the presence of a compact massive torus as well as
the high mass of the PNS, the total neutrino luminosity is enhanced during the evolution
of the system. Furthermore, because of the flattened
geometry of these central objects, the neutrino flux is 
enhanced in the polar region. As a consequence, the 
neutrino heating timescale of the stalled shock becomes shorter than
the timescale of the matter infall in the polar region, 
leading to a bipolar explosion. 

By performing numerical-relativity simulations, we will illustrate that
this mechanism can indeed work for a rapidly rotating progenitor of 
$M_{\rm ZAMS}\approx 
45$--$65M_\odot$, which corresponds to the range of He-core mass of $M_\mathrm{He}=20$--$32M_\odot$. For such high-mass rapidly-rotating progenitors, 
the total mass of the PNS
and surrounding torus becomes also high, and hence, the neutrino luminosity
is enhanced as well. As a result, the bipolar outflow becomes more
energetic than the ordinary SNe. 
Thus, this mechanism may produce a class of energetic SNe like
broad-line type-Ic SNe (see, e.g.,
\citet{Woosley2006a,Cano2017a} for reviews).

This article is organized as follows. 
In \S~\ref{sec:method}, we
summarize the progenitor models employed as the initial condition for
numerical-relativity simulations together with a brief summary of our
method for the simulation.  The results for the successful explosion
are shown in \S~\ref{sec:results}.  Section~\ref{sec:summary} is
devoted to a summary and discussions.

\section{Models and Method} \label{sec:method}

\begin{table*}[t]
\caption{List of the models and the results.
$t_\mathrm{exp}$ and $t_\mathrm{BH}$ denote the post-bounce time at the onset of the explosion and that of the BH formation, respectively.
}
\begin{center}
\begin{tabular}{lccccccc}

\hline \hline

Model & $M_\mathrm{He}$ & $\Omega_0$ & $R_0$ & $(\chi_{5M_\odot},\chi_{10M_\odot}$) & $t_\mathrm{exp}$ & $t_\mathrm{BH}$ & $E_\mathrm{exp}$ \\
 & $(M_\odot)$ & (rad/s) & (km) &  & (s) & (s) & ($10^{51}$\,erg) \\
\hline
M20-0     & 20 &  0  & ---  & (0,0) & --- & 0.3 & --- \\
M20-S040  & 20 & 0.40& 6000 & (1.0, 0.34) & --- & 1.1 & --- \\
M20-S050  & 20 & 0.50& 6000 & (1.3, 0.43) & 3.3 & 4.3 & \addsf{4.2} \\
M20-S075  & 20 & 0.75& 6000 & (1.9, 0.65) & 4.8 & 7.2 & \addsf{4.5} \\
M20-S100  & 20 & 1.00& 6000 & (2.5, 0.87) & 5.9 & 9.8 & \addsf{3.6} \\
M20-L050  & 20 & 0.50& 8500 & (2.2, 1.1) & 7.4 & 9.1 & \addsf{1.2} \\
M20-S050N & 20 & 0.50& 6000 & (1.3, 0.43) & 3.5 & 4.3 & \addsf{1.7} \\
\hline
M32-0     & 32 & 0   & ---  & (0,0) & --- & 0.1 & ---  \\
M32-S050  & 32 & 0.50& 5800 & (1.1, 0.61) & --- & 1.0 & ---  \\
M32-S075  & 32 & 0.75& 5800 & (1.7, 0.92) & 2.7 & 4.3 & \addsf{52} \\
M32-S100  & 32 & 1.00& 5800 & (2.3, 1.2) & 3.2 & 5.1 & \addsf{26} \\
M32-S075DD2 & 32 & 0.75& 5800 & (1.7, 0.92) & 2.8 & 4.4 & \addsf{58} \\
M32-S075N   & 32 & 0.75& 5800 & (1.7, 0.92) & 3.0 & 4.3 & \addsf{11} \\
M32-S075-modE & 32 & 0.75& 5800 & (1.7, 0.92) & 2.6 & 4.3 & \addsf{66} \\

\hline
\end{tabular}
\end{center}
\label{tab:model}
\end{table*}

We employ the final state of high-mass stellar evolution models as the initial condition of our numerical-relativity simulations. 
The stellar evolution of non-rotating He-star models with their initial mass of $M_{\rm He}=20$ and $32M_\odot$ is calculated using the code described in~\citet{Takahashi2018}. 
For these models, $M_{\rm ZAMS}\approx 45$ and $65M_\odot$, respectively~\citep{Woosley2002a}. The evolution calculation is performed until the central temperature reaches $\approx 8 \times 10^9$ K. At this stage, the central density is $\approx 8 \times 10^8\,{\rm g/cm^3}$ for $M_{\rm He}=20M_\odot$  and $\approx 5 \times 10^8\,{\rm g/cm^3}$ for $M_{\rm He}=32M_\odot$.

Observationally, it is known that at least some broad-line type-Ic SNe are associated
with gamma-ray bursts \citep[GRBs;][]{Cano2017a}, and 
theoretically, a broadly accepted candidate for the central engine of GRBs 
is a system composed of a rapidly spinning BH and a dense accretion torus. 
For the formation of the BH-torus system, a rapidly rotating  
progenitor star is obviously necessary~\citep{Woosley1993,Macfadyen1999}.
Hence, we consider rapidly rotating massive stars as the progenitor of such energetic SNe.
Rapidly rotating progenitors may be formed via peculiar single-star evolution~\citep{Yoon2005,Woosley2006} or by binary merger~\citep[e.g.][]{Fryer2005}.
Recent stellar-evolution simulations predict that cores of the fastest rotating core-collapse progenitors may have $\approx 3 \times 10^{16}$\,cm$^2$\,s$^{-1}$ of the averaged specific angular momentum, or equivalently, $\approx 1.36$ of the spin parameters,  inside the enclosed mass of 5 $M_\odot$~\citep{Aguilera-Dena2018}.
These results should be interpreted with caution, however, as even the most advanced stellar-evolution simulations take into account the effects of angular-momentum transport via the convection, circulation, and magnetohydrodynamics (MHD) in a phenomenological manner.

The other point to be stressed is that during the long-term evolution of the PNS and the torus surrounding it, which are the typical outcomes for the rapidly rotating stellar core collapse, the angular momentum transport can play an important role for the evolution of the system, because the timescale of viscous angular momentum transport, which is likely to stem from MHD instabilities, is typically several hundred milliseconds \citep{fujibayashi2018a,Fujibayashi2020a,Fujibayashi2020b}, while we follow the evolution of the system for seconds.
It is also possible that MHD effects such as magnetic braking play an important role for the angular momentum redistribution. Thus, the specific angular momentum distribution of the progenitor star is likely to be significantly modified during the long-term evolution of the system. However, currently, it is not clear how efficiently such angular momentum redistribution proceeds. 

Thus in this paper, as a first step toward a more detailed study, we add an ad-hoc, simple angular momentum profile to the final state of the evolved stars for the initial conditions, with which a system composed of a central object (either an NS or BH) surrounded by a massive torus is formed, even in the absence of the angular momentum
redistribution during the evolution of the collapse outcome (note that no angular momentum 
transport effect is taken into account in this work).

Specifically, the following cylindrical profile is imposed for the angular velocity:
\begin{align}
\Omega = \Omega_0 e^{-R^2/R_0^2}, \label{eq:Omega}
\end{align}
where $\Omega_0$ is the angular velocity along the rotation axis
($z$-axis), $R$ the cylindrical radius, and $R_0$ a cut-off radius.
This rotational profile is somewhat different from the one obtained in the 
stellar-evolution simulations, in which the angular velocity is described as 
a function of the radius in the spherical polar coordinates.
However, because the contribution of the matter along the rotation axis to the mass and angular momentum of the star is minor, the effect of the difference in the profile from spherically symmetric one is likely to be minor.

For $R_0$, we choose the radius at the edge of the Si layer (L model), at which the entropy profile becomes discontinuous, or 70\% of this radius (S model).
Equation~(\ref{eq:Omega}) implies that for $R \ll R_0$, the progenitor star is approximately rigidly rotating, while for the outer region, 
stellar matter rotates slowly.
Such a state is reasonable if the efficiency of the angular momentum transport in the compact central region is sufficiently high.
The steep cut-off of the angular velocity is achieved during the stellar evolution in the presence of the convective layer associated with the shell burning, in which the angular momentum at the bottom of the layer is transported to a large radius.

Table~\ref{tab:model} lists the models considered in this work.
M20 and M32 denote the models with $M_{\rm He}=20$ and $32M_\odot$, respectively. 
The letters ``S" and ``L" refer to the choice of $R_0$ and the following three-digit numbers denote the value of $\Omega_0$ in units of 0.01\,rad/s.
We also perform simulations omitting the neutrino pair-annihilation heating 
(models M32-S075N and M20-S050N) 
to show that this effect contributes substantially to increasing the explosion energy.
To indicate the rapidness of the stellar rotation, 
in Table~\ref{tab:model}, we present a dimensionless spin parameter 
defined by $\chi_M=cJ/GM^2$ where $J$ and $M$ are total angular momentum and 
rest mass enclosed in mass shells at $M=5M_\odot$ and $10M_\odot$. 
\addkt{We note that the values of $\chi_M$ are broadly comparable to the results of a state-of-the-art stellar evolution simulation \citep[e.g.,][]{Aguilera-Dena2018}.}

A finite-temperature EOS referred to as SFHo~\citep{steiner2013a} is employed in this work except for model M32-S075DD2, in which another EOS referred to as DD2 \citep{banik2014a} is employed for comparison.
With SFHo and DD2 EOSs, the maximum values of the gravitational mass for 
the non-rotating cold NSs are $M_\mathrm{max} \approx 2.06$ and 2.42$M_\odot$, and the radii of the non-rotating NSs with mass $1.4M_\odot$ are 11.9 and 13.2\,km, respectively.
The SFHo EOS is relatively soft in the sense that the value of $M_\mathrm{max}$ is close to $2M_\odot$ and the radius is relatively small as $\alt 12$\,km.

With the setting listed in Table~\ref{tab:model}, the PNS formed after the collapse is rapidly
rotating and the resulting centrifugal force plays an important role
to allow the rest mass of the PNS beyond $3M_\odot$ (cf.~Fig.~\ref{fig:fig1}
in \S~\ref{sec:results}). 
We note that several other simulations, for which we do not present the results 
in this article, already confirmed the collapse to a BH without the shock revival for $\Omega_0 \leq 0.4\,\mathrm{rad/s}$ for $M_\mathrm{He}=20M_\odot$ and for $\Omega_0 \leq 0.5\,\mathrm{rad/s}$ for $M_\mathrm{He}=32M_\odot$.
We also note that, for non-rotating models, the PNS collapses into a BH approximately at 0.1\,s and 0.3\,s after bounce for $M_\mathrm{He}=32M_\odot$ and $20M_\odot$ models, respectively.

Numerical-relativity simulations are performed with our latest axisymmetric neutrino-radiation viscous-hydrodynamics code.
The details are described in~\citet{fujibayashi2017a,Fujibayashi2020b}.
In this work, we do not take the viscosity into account.

Einstein's equation is solved with the original version of the puncture-Baumgarte-Shapiro-Shibata-Nakamura formalism \citep{shibata1995a,baumgarte1999,marronetti2008} incorporating the Z4c prescription \citep{Hilditch2013a} for the constraint violation propagation.
We solve geometrical variables in Cartesian coordinates and employ the so-called cartoon method to impose axisymmetry to them \citep{Shibata2000a,Alcubierre2001a}. The spatial 
interpolation necessary for the cartoon process is carried out using the 
fourth-order accurate Lagrangian interpolation. 

The radiation hydrodynamics equations are solved with a version of the leakage scheme together with a moment transport scheme.
The detailed description of the schemes is found in \cite{sekiguchi2010a} and \cite{fujibayashi2017a}.
In this method, the emitted neutrinos are divided into ``trapped" and ``streaming" components.
The trapped neutrinos are assumed to be thermalized with the fluid, and treated as a part 
of the fluid. In our numerical scheme, 
they are diffused out to the streaming component in the diffusion timescale.

The streaming neutrinos are solved using energy-integrated truncated-moment formalism \citep{shibata2011a} with the so-called M1-closure relation to estimate the higher moments.\footnote{It is well-known that the crossing multiple beams cannot be appropriately solved with moment-based schemes, and it can be a source of the systematic error on the neutrino distribution and heating rate in the system with a non-spherical hydrodynamical profile.
In~\citet{Sumiyoshi2021}, the Eddington tensor is evaluated with a Boltzmann neutrino transfer code and compared with that by an M1-closure relation for a system composed of a massive NS and a torus formed in a binary NS merger.
It is found that the deviation of the Eddington tensor from that derived with the M1-closure is at most 10\% in the edge of the NS and torus, and in the polar region the deviation is smaller.
The system that we consider in this work has a similar profile to the one investigated in~\citet{Sumiyoshi2021}, and hence, we expect that the systematic error associated with the moment-based scheme could not be significant, although for more quantitative studies a better radiation transfer scheme such as those in \citet{Harada2020} and \citet{Foucart2020} is obviously required. 
}
The heating due to neutrino absorption and pair-annihilation is included in an approximate manner~\citep{fujibayashi2017a}.

The free parameters of our leakage scheme \citep[see][for the parameters]{sekiguchi2010a} are phenomenologically determined. Specifically, we performed simulations for the collapse of a 15\,$M_\odot$ solar-metallicity progenitor \citep{Woosley2002a} and compared the neutrino luminosity at the core bounce with those  in~\cite{Liebendoerfer2003a} and \cite{Janka2012b}.
We then employ the parameters by which their neutrino luminosity curves are approximately reproduced.
The heating of the matter by streaming neutrinos is conservatively incorporated in this work: specifically, the heating term is reduced by a factor of $\exp(-2\tau_i)$ with the optical depth of $i$-th species of neutrinos $\tau_i$ (i.e., this factor is multiplied to the opacity). 
Thus, only for sufficiently outside the neutrino spheres, the heating is efficient. 

In energy-integrated neutrino transfer schemes, the heating rate due to the neutrino-matter interaction depends on the method for estimating neutrino energy distribution through the energy dependence of the neutrino cross section \citep[see, e.g., ][]{Foucart2016b}. 
To illustrate this dependence, we perform an additional simulation with the same setup as M32-S075, but with a different method for its estimation (see Appendix~\ref{app:energy} for the detail and results).

For numerical simulations, we employ the same nonuniform grid as that in our previous work \citep{Fujibayashi2020b}.
In the inner region with $z<15$\,km and $R<15$\,km, the uniform grid is prepared with the grid spacing of 150\,m.
In the outer region, the nonuniform grid is prepared with the increase rate of the grid spacing of 1.01.
The computational domain is $ 0 \leq x \leq L$ and $ 0 \leq z \leq L$ with $L\approx 3\times 10^4$\,km.

\section{Simulations Results} \label{sec:results}

\begin{figure}[t]
\includegraphics[width=0.48\textwidth]{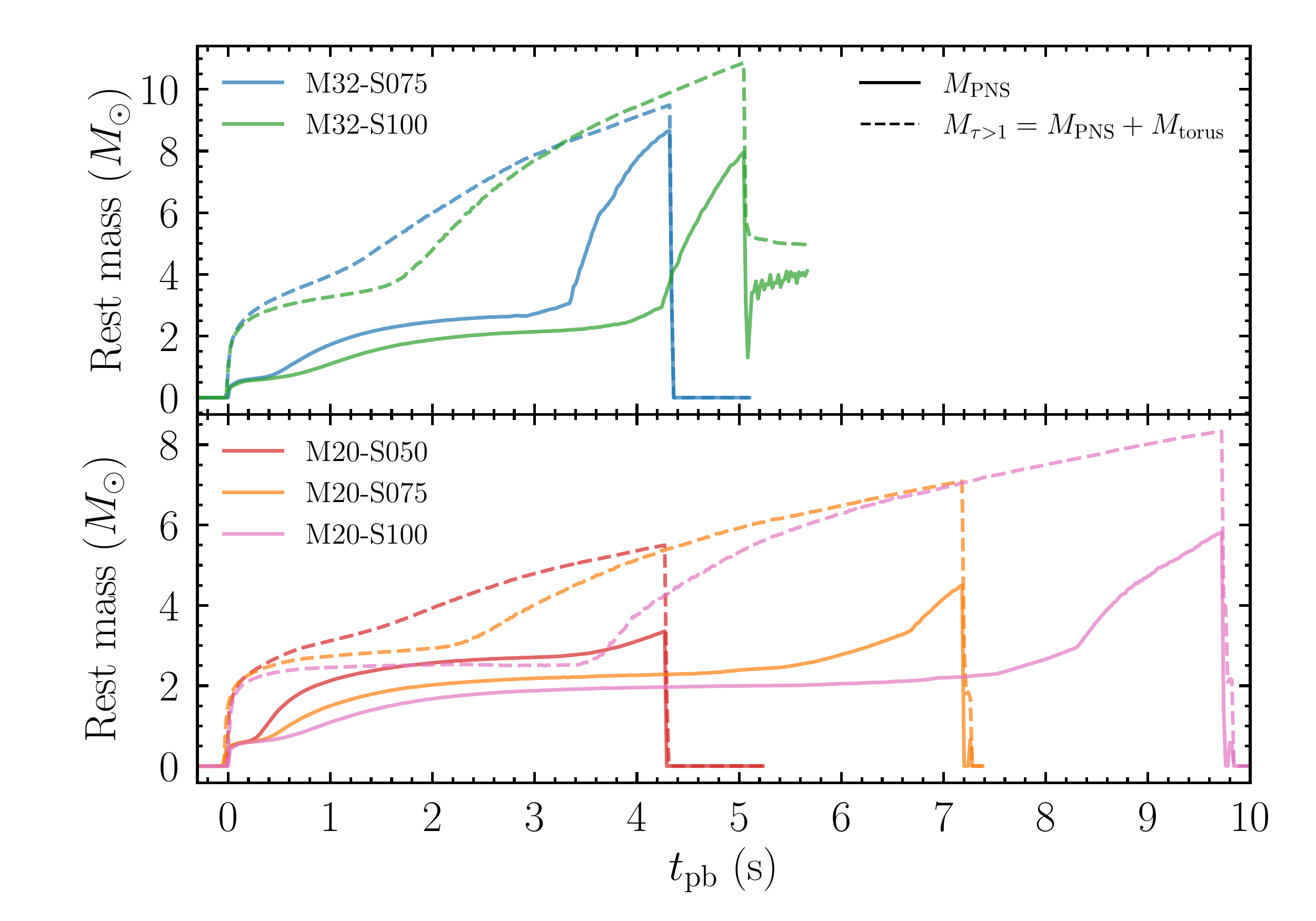}
\caption{Evolution of the rest mass of the PNS (solid curves) and that 
in the optically thick region for neutrinos (dashed curves) for selected models.
\textbf{Note.} Due to the definition of it, the contribution from the torus is 
included in the value of $M_\mathrm{PNS}$, and thus, it becomes 
very large just prior to the collapse to the BH. 
For the same reason, $M_\mathrm{PNS}$ has a finite value even after the BH formation for model M32-S100.
}
\label{fig:fig1}
\end{figure}

\begin{figure*}[t]
\includegraphics[width=\textwidth]{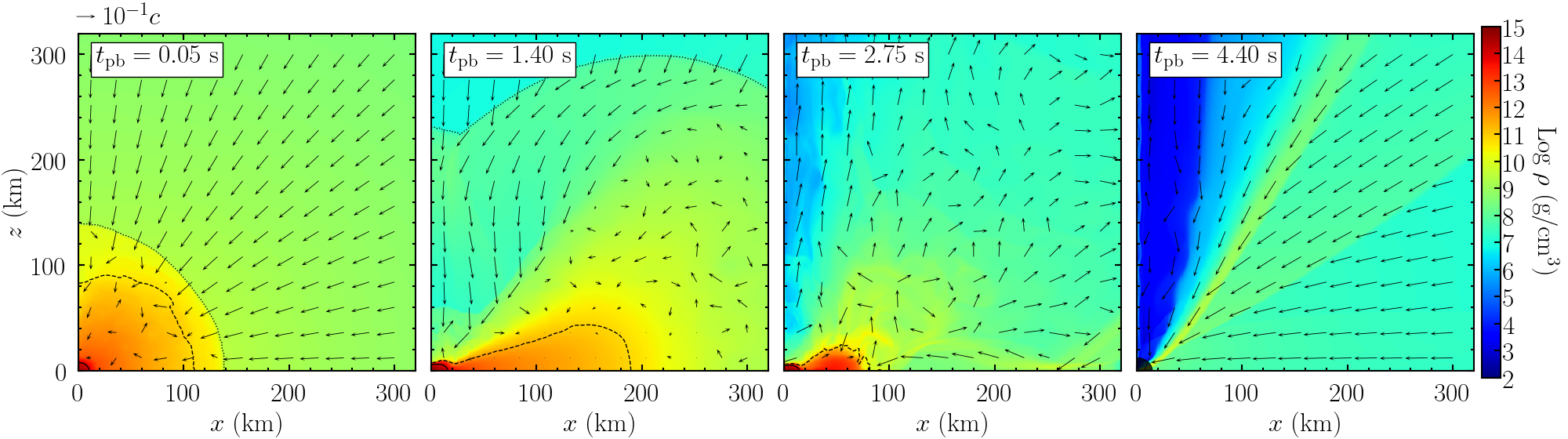}
\caption{
Snapshots of the rest-mass density at $t_\mathrm{pb}=0.05$, 1.40, 2.75, and 4.40\,s for model M32-S075.
The solid, dashed, and dotted curves denote the surfaces with the density of $\rho=10^{14}$\,g/cm$^3$, with the neutrino optical depth $\tau=1$, and at the stalled shock, respectively.
In the fourth panel, a BH is formed at the center (shaded region).
On all the panels, the arrows display the poloidal velocity field ($v^x,v^z$). 
Their length is proportional logarithmically to the velocity. 
See also an animation at
\url{http://www2.yukawa.kyoto-u.ac.jp/~sho.fujibayashi/share/anim_den_M32-S075.mp4}.}

\label{fig:fig2}
\end{figure*}

For all the simulations, a PNS is first formed after the stellar-core collapse.
Then, the baryon rest mass of the PNS increases to $M_\mathrm{PNS}= 2.0$--$2.5M_\odot$ in $t_\mathrm{pb}:=t-t_\mathrm{b} \sim 1$\,s
(see~Fig.~\ref{fig:fig1}).
Here $t_\mathrm{b}$ denotes the time at the core bounce and we defined $M_\mathrm{PNS}$ to be the total rest mass in the region of $\rho \geq 10^{14}\,\mathrm{g/cm^3}$ 
(we note that due to this definition, a part of the rest mass of the dense region of the 
torus is included 
in $M_\mathrm{PNS}$ just prior to the BH formation). 
Subsequently, $M_\mathrm{PNS}$ exceeds $3M_\odot$ for all the rapidly rotating models listed in Table~\ref{tab:model}. 
This mass exceeds the maximum rest mass of the non-rotating cold NSs, which is $\approx 2.42 M_\odot$ and $\approx 2.89M_\odot$ for the SFHo and DD2 EOSs, respectively. 
Thus, the centrifugal force (and partly the thermal pressure) plays a key role for preventing the collapse of the PNS to a BH for seconds. 
Along the rotation axis the rotational period becomes $\sim 0.5$\,ms in the late stage of the PNS.  

Together with the PNS, a massive torus is formed around it.
Here, we define the torus mass by $M_\mathrm{torus}:=M_{\tau>1}-M_\mathrm{PNS}$, where $M_{\tau>1}$ is the total rest mass in a region with the average optical depth of electron neutrinos and antineutrinos ($\tau\equiv (\tau_{\nu_\mathrm{e}}+\tau_{\bar{\nu}_\mathrm{e}})/2$) larger than unity. 
We find that the torus mass increases by the matter infall and eventually far exceeds $1M_\odot$.
For $M_\mathrm{He}=32M_\odot$ models, this mass becomes very large in a short post-bounce time.
The torus initially has a radius of $\sim 200$\,km on the equatorial plane (see the second panel of Fig.~\ref{fig:fig2} for M32-S075; the dashed curve).
During the growth of the torus, a standing accretion shock with a donuts shape is formed surrounding the PNS and torus (the second panel of Fig.~\ref{fig:fig2}; the dotted curve),  and this shock expands gradually with time due to the shock heating induced by the matter infall.
Because of our choice of the initial angular-velocity profile, the matter that accretes onto the PNS and torus at late times has smaller specific angular momenta.
Because of its high mass and less specific centrifugal force at late times, the torus shrinks (its density increases; the third panel of Fig.~\ref{fig:fig2}), and as a result, the value of $M_\mathrm{PNS}$ steeply increases prior to the formation of a BH (see the upper panel of Fig~\ref{fig:fig1}). For the larger value of $R_0$ for which the specific angular 
momentum of the matter in the outer region is larger, the mass accretion timescale is longer.

\begin{figure}[t]
\includegraphics[width=0.48\textwidth]{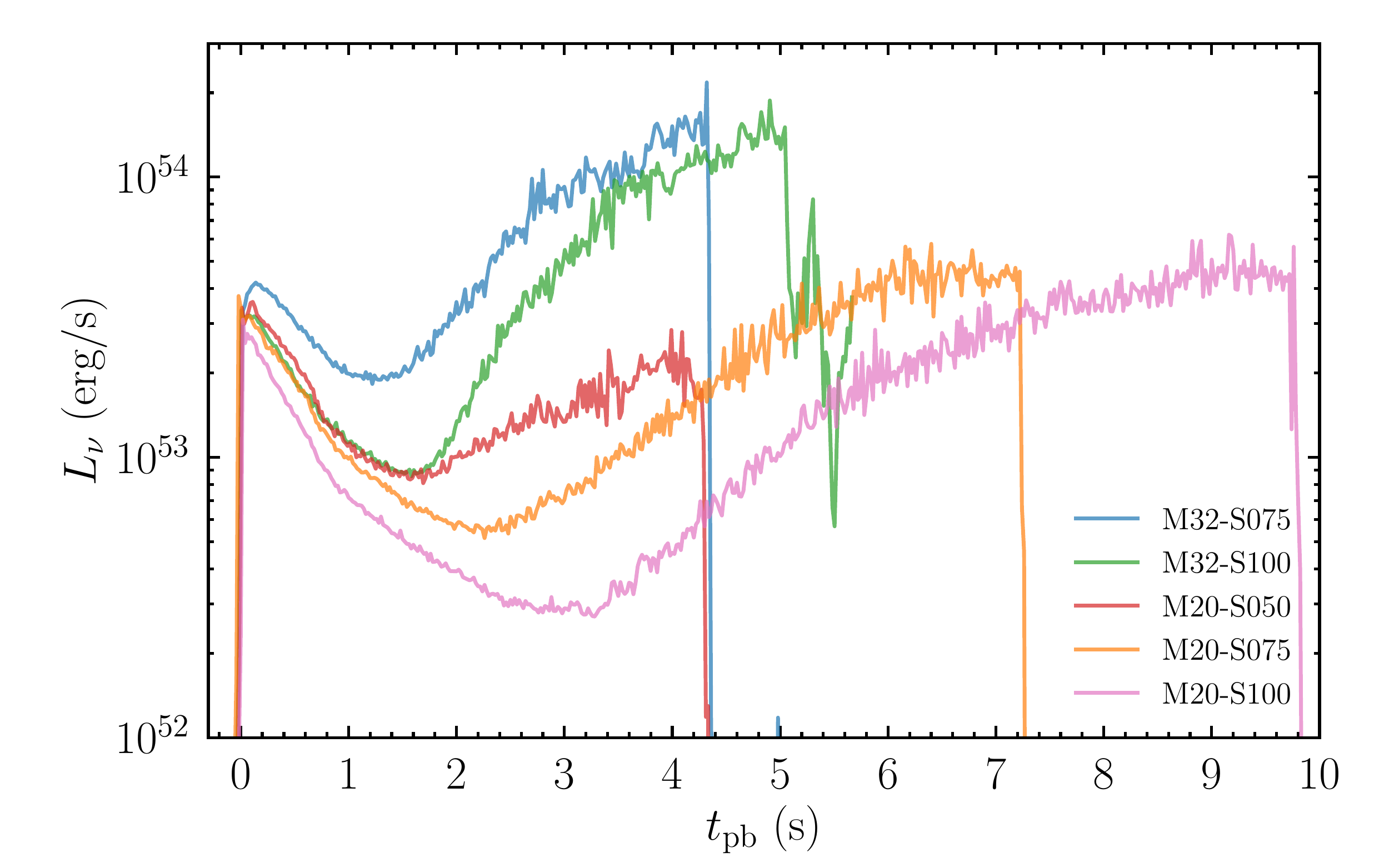}
\includegraphics[width=0.48\textwidth]{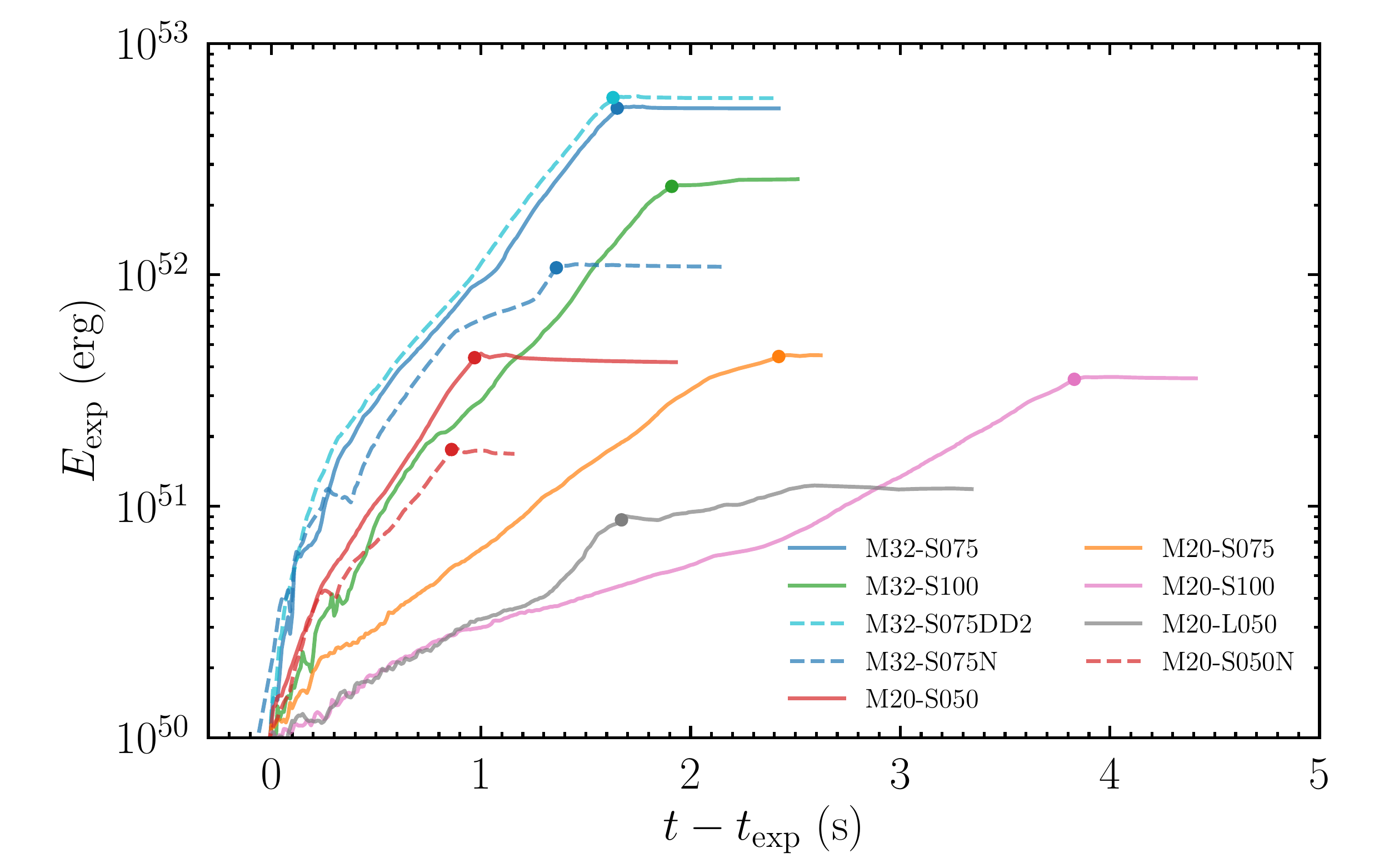}
\caption{Top: Total neutrino luminosity as a function of post-bounce time. Bottom: Diagnostic explosion energy as a function of $t-t_\mathrm{exp}$, where $t_\mathrm{exp}$ is the explosion time defined as the time at which the explosion energy exceeds $10^{50}$\,erg.
The filled circle on each curve denotes the BH formation time for each model.
}
\label{fig:fig3}
\end{figure}

\begin{figure}[t]
\includegraphics[width=0.48\textwidth]{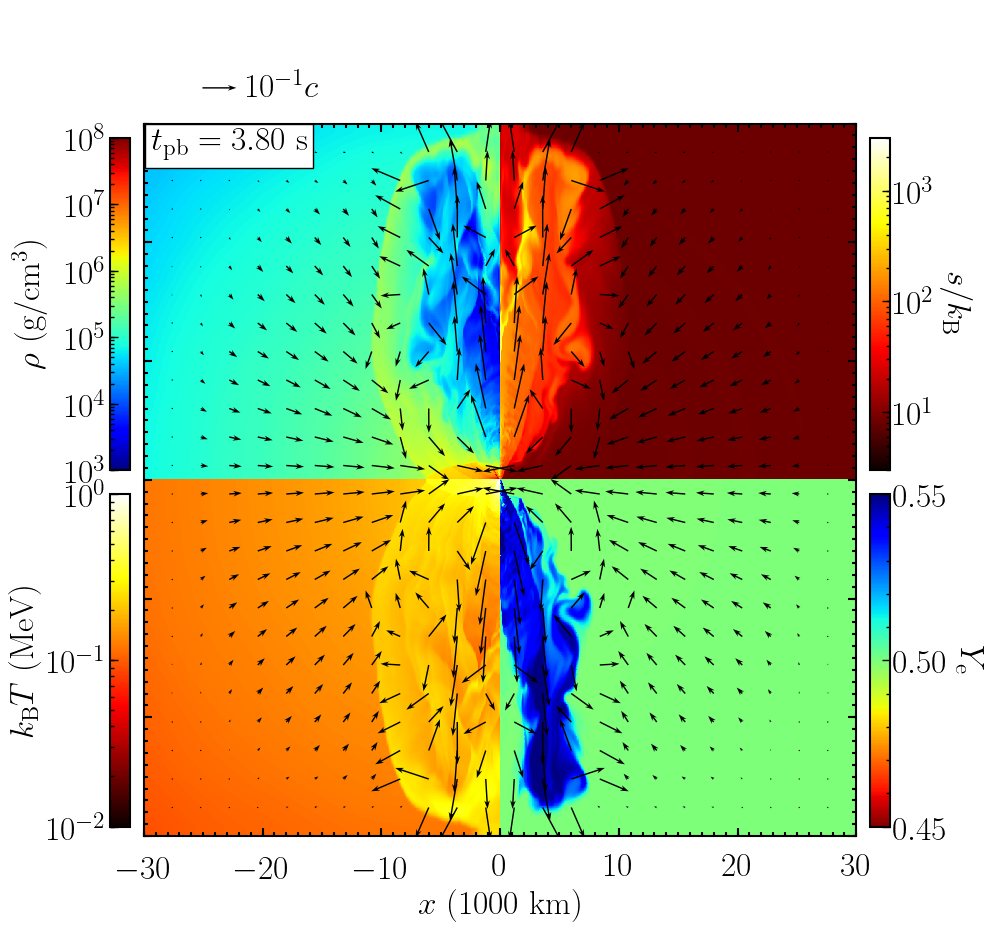}
\caption{Snapshots of the rest-mass density (top-left), entropy per baryon (top-right), temperature (bottom-left), and electron fraction (bottom-right) at $t_\mathrm{pb}=3.80$\,s for model M32-S075. \addms{$k_{\rm B}$ denotes the Boltzmann's constant.}
The arrows display the poloidal velocity field ($v^x,v^z$). Their length is proportional logarithmically to the velocity. See also an animation at
\url{http://www2.yukawa.kyoto-u.ac.jp/~sho.fujibayashi/share/anim_4_M32-S075_30000.mp4}.
}
\label{fig:fig25}
\end{figure}

The shrink of the torus enhances the neutrino luminosity (see Fig.~\ref{fig:fig3} for the increase of it in late stages), in particular from the torus.
The maximum neutrino luminosity is higher for the higher values of $M_\mathrm{He}$ and could be close to $10^{54}\,$erg/s as found in~\citet{Sekiguchi2011}. 
Because the ram pressure by the infalling matter decreases with time, such huge neutrino heating naturally leads to the shock revival.
The explosion occurs in particular toward the polar direction for which the matter density and associated ram pressure are relatively small (see Fig.~\ref{fig:fig25}). 
The explosion occurs qualitatively in the same manner for all the rapidly rotating models listed in Table~\ref{tab:model}.

Table~\ref{tab:model} lists the diagnostic explosion energy, $E_\mathrm{exp}$. % kinetic energy of the ejecta for the bipolar explosion, $E_\mathrm{kin}$.
Here, this explosion energy is evaluated in the computational region of $\alt 30000$\,km by integrating the positive binding energy of the matter as in \citet{Mueller2012} (see Appendix A for our formulation to it).
For the present explosion models, $E_\mathrm{exp}$ eventually exceeds $10^{51}\,{\rm erg}$, and for $M_\mathrm{He}=32M_\odot$ models, it becomes higher than $10^{52}$\,erg, i.e., appreciably higher than the kinetic energy of typical SNe.
This is the reflection of the neutrino luminosity by one order of magnitude higher than in the typical SNe \citep[on the relation between the neutrino luminosity and explosion energy, see, e.g.,][]{Yamamoto2013}.
\addsf{
The explosion energy for models without the neutrino-antineutrino pair annihilation process (M20-S050N and M32-S075N) is 2.5--4 times smaller than that for corresponding models with the process (see Table~\ref{tab:model}).
This indicates that the pair-annihilation of neutrinos is the dominant process of the energy injection.
}

This lager neutrino-driven energy injection could be a substantial fraction of the energy injection for broad-line type-Ic SNe with a bipolar outflow~\citep{Maeda2002a,Maeda2003a,Mazzali2005a,Maeda2008a}.
We note that the energy deposition rate to the outflow is $\gtrsim 10^{52}$\,erg/s for $M_\mathrm{He}=32M_\odot$ models.
Thus, $^{56}$Ni with mass of $10^{-2}$--$10^{-1}M_\odot$ may be synthesized in the ejecta~\citep[][Wanajo et al.~in preparation]{Tominaga2007}.

For model M32-S075DD2, the explosion energy is slightly higher than that for M32-S075, but the difference is not significant.
This is because the lifetime of the PNS and the duration of the energy injection through the neutrino heating are only slightly different between the two models with different EOSs due to the rapid increase of $M_\mathrm{PNS}$ to the critical mass for the gravitational collapse to a BH.

\addms{For the larger value of $R_0$, the explosion is delayed and the explosion energy is smaller (compare the results of M20-S050 and M20-L050). The reason for this is that for the larger value of $R_0$ (for a given value of $\Omega_0$), the increase of the torus mass is delayed due to the larger centrifugal force, and the specific neutrino emissivity is decreased in the late time at which the explosion is driven (i.e., the formation timescale of the torus is as long as or longer than the neutrino cooling one). 
This suggests that the angular momentum distribution is the key for controlling the explosion energy.}

We also note that for model M32-S100, the massive torus remains, maintaining a high neutrino luminosity ($\gtrsim 10^{52}$\,erg/s), after the BH formation (see Figs.~\ref{fig:fig1} and \ref{fig:fig3}).
As a result, the neutrino-driven outflow is still present after the BH formation.
However, the neutrino luminosity is not enhanced significantly and the explosion energy is relatively low for this model, in spite of the formation of a massive torus.
The reason for this is, \addms{again}, that the formation timescale of the torus is as long as or longer than its neutrino cooling one in this model.
Hence, to enhance the neutrino luminosity far beyond $10^{53}$\,erg/s, the torus has to be formed before the sufficient neutrino cooling occurs.
For achieving such a physical state, a pre-collapse progenitor with a compact core, which 
has angular momentum sufficiently large in its inner region with a steep cut-off at a 
radius, is likely to be necessary.

For the high-mass progenitors employed in this work, 
a BH is eventually formed due to the continuous 
matter accretion onto the PNS, in particular from the equatorial direction.
Since the central object gains a large amount of the angular momentum from the rotating progenitor, the BH at the formation is rapidly spinning with the dimensionless spin $\gtrsim 0.9$ irrespective of the models.
The rest-mass density in the vicinity of the rotation axis becomes low as $\alt 10^3\,{\rm g/cm^3}$ after the BH formation (see the fourth panel of Fig.~\ref{fig:fig2}). 
The eventual total rest mass in the funnel region of 
$R \leq 50$\,km and $|z| \leq 10^4$\,km is 
$\alt 10^{-7}M_\odot$.

\section{Summary and Discussions} \label{sec:summary}
This article proposes a new mechanism for driving energetic SNe like broad-line type-Ic SNe by the neutrino heating.  The
model supposes that the progenitor stars of the SNe have high mass and
rotation rapid enough to form a rapidly rotating PNS surrounded by a
high-mass torus. The resulting PNS can survive for seconds due to the
strong centrifugal-force support, and in addition, due to the presence of a
high-mass torus surrounding it, the total neutrino luminosity can be
quite high at several seconds after the core bounce. Then, the
neutrino heating drives a high-energy SNe, in particular toward the polar direction.  In the successful explosion, the total rest mass of the central object becomes high enough (i.e.,
$\agt 3M_\odot$) and the explosion is significantly delayed, typically
for seconds, after the core bounce (thus to study this model, we need a long-term
general relativistic simulation, only by which the criterion for the 
formation of the BH is accurately taken into account). In this case, the explosion occurs 
in a bipolar manner via the neutrino heating. 
The explosion energy is beyond the typical explosion energy of SNe $\sim 10^{51}\,{\rm erg/s}$, and even larger than $10^{52}$\,erg for high-mass progenitor models.
Thus, this mechanism could provide (at least a part of) energy-injection for energetic SNe like broad-line type-Ic SNe.

Furthermore, a rapidly spinning BH is eventually formed. 
Since the polar outflow found in this paper produces a low-density funnel along 
the rotation axis, the remnant looks suitable for launching an ultra-relativistic jet, i.e., GRB~\citep[][]{Woosley1993,MacFadyen2001,Woosley2006a,Cano2017a}, in the presence of 
an energy injection. 
We here note that the mechanism for launching the ultra-relativistic jet is not necessarily the same as that for inducing the bipolar outflow; e.g., an MHD 
process may be the source for GRBs; see, e.g.,~\citet{Piran2004}.
If the formed BH is surrounded by a magnetized massive torus, such a system could drive a relativistic jet by subtracting the rotational kinetic energy of the BH~\citep{Blandford1977}.
The relativistic jet could not only drive a GRB but also be the additional energy injection for the SN explosion. Thus, this model provides a scenario for the association of broad-line type-Ic SNe and GRBs.

However, in the angular-velocity profiles chosen in this paper, 
the matter initially located in large radii has small angular momenta.
Thus, except for M20-L050, the torus mass is not very large after the BH formation, and hence, 
in the present models, 
it is unlikely to subsequently cause long-term energetic phenomena powered by the accretion of the torus matter onto a BH.
By contrast, if the matter in the outer region initially has larger specific angular momenta than that in the central region, which may be a reasonable assumption considering more realistic stellar evolution, a massive accretion torus can be formed after the BH formation.
In such a case, a further activity of the system is expected. 
Exploring this possibility is interesting future work.

The above speculation suggests that the presence or absence of the activity after the bipolar explosion may depend on the angular momentum distribution of the progenitor stars, and this may explain a variety of the activity 
duration of the central engine and 
a variety of the high-energy events associated with the 
broad-line type Ic SNe~\citep{Woosley2006,Margutti2014,Lazzati2012} (see also~\citet{Nakar2015} on the importance of the density profile of the pre-collapse progenitor). 
Our numerical results also match with the speculation that 
the rapidly rotating massive stars are likely to be 
the progenitors for the energetic type-Ic SNe and GRBs~\citep{Yoon2005,Fryer2005,Woosley2006,Aguilera-Dena2018}. 

Recent radiation-MHD simulations (in non-relativistic gravity) by \citet{Obergaulinger2017,Obergaulinger2020,Obergaulinger2021,Aloy2021a} also have shown that in the presence of a rapid rotation, a high-mass progenitor star can explode by the combination of the neutrino heating, rotation, and magnetic-field effects.
Our result is similar to theirs, but our work shows that an energetic explosion can occur purely by the neutrino heating effect even in the absence of magnetorotational effects, for the progenitor stars more massive than that employed in \citet{Obergaulinger2017,Obergaulinger2020,Obergaulinger2021,Aloy2021a}.
The only required condition for our case is the presence of sufficiently rapid rotation 
inside the stellar core. 

In this paper, we present only models that show the explosion.
For low angular-momentum models, the PNS collapses to a BH before the explosion.
Thus, for the explosion, the progenitor stars need to have a sufficient angular momentum.
The details on the non-explosion models and approximate criterion for the explosion should be systematically studied. 

There are several issues to quantitatively improve the present work. 
First, our treatment for the neutrino-radiation transfer is currently based on a gray leakage scheme. Obviously, simulations with a better radiation transfer code are needed.

The present work is based on axisymmetric simulations. Because the torus is massive,
non-axisymmetric deformation is likely to take place in reality \citep[e.g.,][] {Shibata2005,Shibagaki2020}. This may cause an 
angular momentum transport in the torus and the accretion onto the PNS
may be enhanced leading to earlier collapse to a BH. The
angular momentum transport can also be enhanced by
MHD effects such as the magneto-rotational
instability~\citep{Moesta2014} and magnetic braking. 
Alternatively, MHD effects may help earlier explosion if
the magnetic field is amplified significantly by MHD instabilities~\citep{Obergaulinger2020,Obergaulinger2021}. 
All these possibilities suggest that we need more sophisticated simulations. 
Thus, we plan to investigate the MHD effects using a radiation-MHD 
code recently developed~\citep{Shibata2021a}. 

The non-axisymmetric deformation of the massive torus could also lead to the 
burst emission of gravitational waves. Our latest study shows that if an 
one-armed spiral deformation mode grows in a dynamical timescale comparable 
to the typical rotational period of the torus, the degree of the 
non-axisymmetric density fluctuation can be 10--20\% of the torus mass~\citep{Shibata2021}. 
In such deformation, the maximum amplitude of burst-type gravitational waves 
at the hypothetical distance to the source of 100\,Mpc can be 
$\sim 10^{-22}$ with the typical frequency of 0.7--0.8\,kHz for 
$M_\mathrm{PNS}\approx 3M_\odot$ with the comparable torus mass~\citep{Shibata2021}.  
Such gravitational waves are the interesting sources for the third-generation 
gravitational-wave detectors such as Einstein Telescope \citep{Punturo2010} and Cosmic Explorer \citep{Abbott2017}. 
Thus, in the future, high-energy supernovae with the bipolar outflow may be 
explored not only by electromagnetic telescopes but also by the 
gravitational-wave detectors.

\acknowledgments

We thank T. Kuroda, K. Maeda, N. Tominaga, and S. Wanajo for useful discussions. 
This work was in part supported by Grant-in-Aid for Scientific
Research (Grant Nos.~JP20H00158) of Japanese MEXT/JSPS.  Numerical
computations were performed on Sakura and Cobra at Max Planck
Computing and Data Facility and XC50 at National Astronomical Observatory of Japan.

\appendix

\section{Diagnosis of the explosion energy} \label{app:energy}

In this Appendix, we describe how to estimate the explosion energy in this work.
The explosion energy is estimated for the matter which are gravitationally unbound and located at a region far from the central object.
In such a far region, the spacetime is approximately stationary and we may consider that an approximately time-like Killing vector exists.
If $(\del_t)^\mu$ is assumed to be the time-like Killing vector, 
the conservation equation of the energy density is described by
\begin{align}
%\del_\mu (\sqrt{-g}T_\nu{}^\mu (\del_t)^\nu) 
\nabla_\mu T^\mu_{~t}
= {1 \over \sqrt{-g}}\del_\mu ( \sqrt{-g} T^\mu_{~t}) = 0,
\end{align}
where $T_{\mu\nu}$ is the energy-momentum tensor of the matter, 
$g$ the determinant of the spacetime metric $g_{\mu\nu}$, and 
$\nabla_\mu$ the covariant derivative with respect to $g_{\mu\nu}$. 
Then, the conserved energy density and associated flux density, respectively, are
defined by 
\begin{align}
 -\sqrt{-g}T_{~t}^t =& - \alpha \sqrt{\gamma}(\rho h u_t u^t + P) = \alpha \sqrt{\gamma}(\rho h w^2 - P -\rho h u^t u_k\beta^k)\notag\\
 =&\ \rho_* (\alpha  \hat{e} - \hat{u}_k\beta^k),\\
 -\sqrt{-g}T_{~t}^i =& - \alpha \sqrt{\gamma} \rho h u_t u^i = \alpha \sqrt{\gamma}(\rho h w^2 v^i - \rho h u^i u_k\beta^k)\notag\\
 =&\  \rho_* v^i (\alpha h w - \hat{u}_k\beta^k),
\end{align}
where $\rho_*=\rho w \sqrt{\gamma}$, $\hat{e}=hw-P/\rho w$, and $\hat{u}_i = hu_i$ with 
the lapse function $\alpha$, the determinent of the spatial metric $\gamma(=g/\alpha)$, 
the pressure $P$, specific enthalpy $h$, and $w=\alpha u^t$.
The specific binding energy $e_\mathrm{bind}$ is then defined by
\begin{align}
e_\mathrm{bind} = \frac{-\sqrt{-g}T_{~t}^t}{\sqrt{-g}\rho u^t} - (1+\varepsilon_\mathrm{min}) = \alpha \hat{e} - \hat{u}_k\beta^k - (1+\varepsilon_\mathrm{min}).
\end{align}
Here, $\varepsilon_\mathrm{min}\approx -0.0013$ is the minimum specific internal energy (including nuclear binding energy) in the employed EOS table.
Note that this definition is slightly different from that in \citet{Mueller2012} due to the presence of the shift vector by $\hat{u}_k\beta^k$.

We define the explosion energy as the volume integral of the positive binding energy density of the matter, i.e, as
\begin{align}
E_\mathrm{exp} =& \int_{e_\mathrm{bind}>0}d^3x \sqrt{-g} (T_{~t}^t-\rho u^t(1+\varepsilon_\mathrm{min})) \notag\\
=& \int_{e_\mathrm{bind}>0,\ r<r_\mathrm{ext}}d^3x \sqrt{-g} T_{~t}^t + \int dt \int_{e_\mathrm{bind}>0,\ r=r_\mathrm{ext}}ds_k \sqrt{-g} T_{~t}^k -(1+\varepsilon_\mathrm{min})M_\mathrm{ej}, \label{eq:Eexp-1}
\end{align}
where $ds_k$ is the area element of a sphere with radius $r_\mathrm{ext}$, and $M_\mathrm{ej}$ is the ejecta mass defined by
\begin{align}
M_\mathrm{ej} =& \int_{e_\mathrm{bind}>0}d^3x \sqrt{-g} \rho u^t = \int_{e_\mathrm{bind}>0,\ r<r_\mathrm{ext}}d^3x \rho_* + \int dt \int_{e_\mathrm{bind}>0,\ r=r_\mathrm{ext}}ds_k \rho_* v^k.
\end{align}

\begin{figure}[t]
\begin{center}
\includegraphics[width=0.48\textwidth]{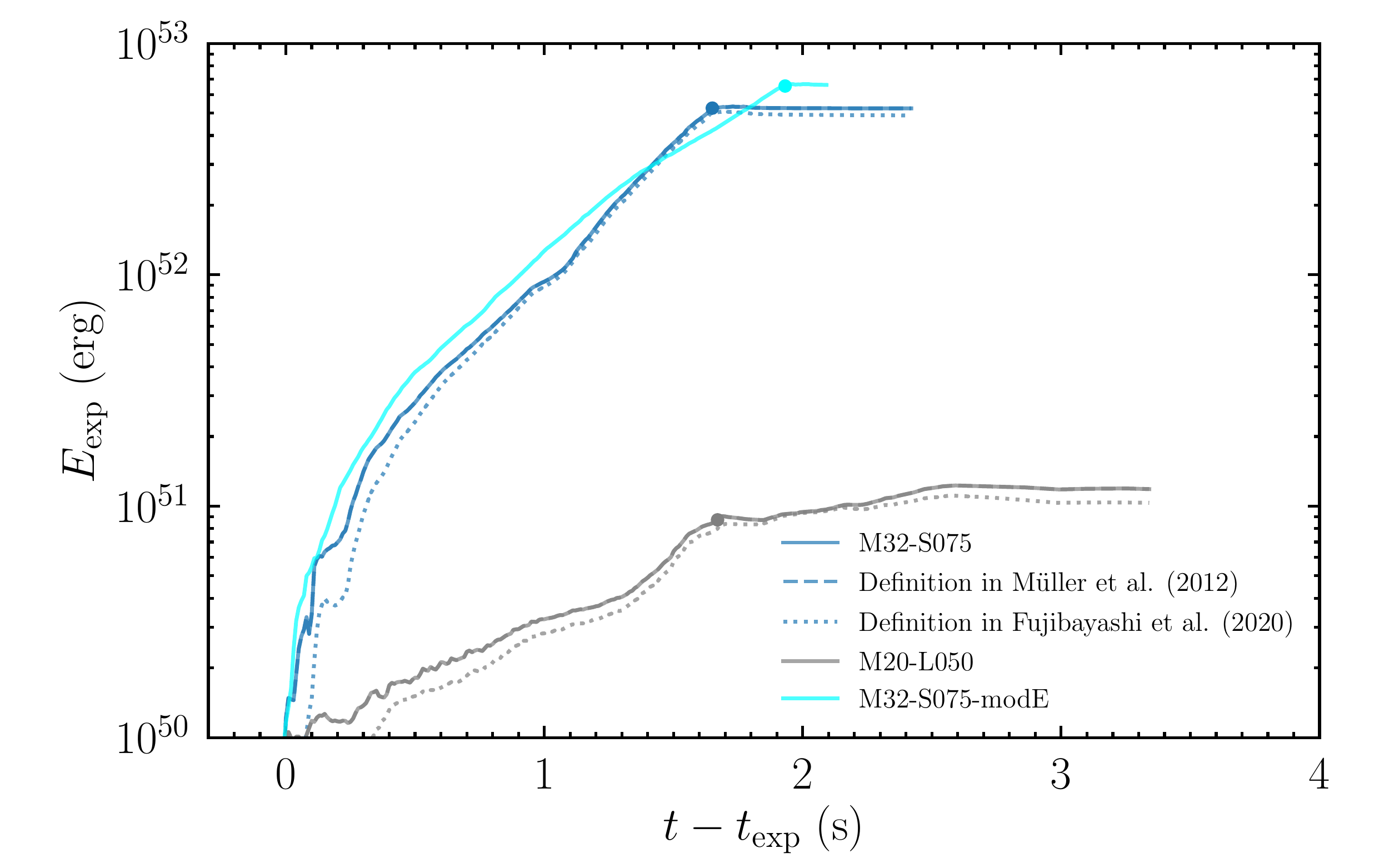}
\caption{Diagnostic explosion energy as a function of $t-t_\mathrm{exp}$ for models M32-S075 and M32-S075-modE.
For model M32-S075, the explosion energy defined by Eq.~\eqref{eq:Eexp-1} (solid), in \cite{Mueller2012} (dashed), and by Eq.~\eqref{eq:Eexp-2} (dotted) are shown.
}
\label{fig:Eexp_app}
\end{center}
\end{figure}

Figure~\ref{fig:Eexp_app} compares the explosion energy defined in this Appendix with 
those by the methods in~\cite{Mueller2012} and in~\cite{Fujibayashi2020a} for models M32-S075 and M20-L050, which have the largest and smallest values of the explosion energy among the models with the SFHo EOS employed, respectively.
Here, in \cite{Fujibayashi2020a}, the ejecta is defined as the matter with $h u_t + h_\mathrm{min} < 0$ based on Bernoulli's argument ($h_\mathrm{min}\approx 1+\varepsilon_\mathrm{min}$ is the minimum specific enthalpy in the employed EOS table),
and the explosion energy is calculated by
\begin{align}
\int_{h u_t + h_\mathrm{min} < 0} d^3x\ \rho_* \biggl( \hat{e} - \frac{M}{r} - h_\mathrm{min}\biggr) &= \int_{h u_t + h_\mathrm{min} < 0,\ r<r_\mathrm{ext}} d^3x \rho_*\biggl(\hat{e}-\frac{M}{r} - h_\mathrm{min}\biggr)\notag\\
&+\int dt \int_{h u_t + h_\mathrm{min} < 0,\ r=r_\mathrm{ext}} ds_k\ \rho_*v^k \biggl(\hat{e}-\frac{M}{r_\mathrm{ext}} - h_\mathrm{min}\biggr), \label{eq:Eexp-2}
\end{align}
where the contribution of the gravitational binding energy is considered by adding $-GM\rho_*/r$ to the total energy density of the ejecta with the gravitational mass of the central object $M$, which is approximated by the enclosed baryon mass at the extraction radius in this case.
Note that, of course, the criteria used in Eqs.~\eqref{eq:Eexp-1} and \eqref{eq:Eexp-2} have the same Newtonian (non-relativistic and weak-gravity) limit.

We find that the methods of estimating the explosion energy in Eq.~\eqref{eq:Eexp-1} and in \cite{Mueller2012} give very similar values. This implies that the contribution of the term with the shift vector ($\hat{u}_k \beta^k$) is negligible in this case.
We also find that the values of the explosion energy defined by Eqs.~\eqref{eq:Eexp-1} and \eqref{eq:Eexp-2} are different only slightly ($\approx 8$\%) for M32-S075.
This is because at the extraction radius ($\approx 3\times 10^4$\,km), the flow is approximately stationary and the Bernoulli's argument gives a good criterion for the ejecta. In addition, the contribution of the gravitational binding energy at the large radius is only $GM/c^2r_\mathrm{ext}\approx 0.05 (M/10M_\odot)$\% of the rest-mass energy, so that it is minor compared to the kinetic energy of the ejecta, and thus, the explosion energy does not depend strongly on the methods of its diagnosis.
For model M20-L050, on the other hand, the difference of the explosion energy is relatively large $\approx 16$\%, likely because the explosion is less energetic and the contribution of the gravitational binding energy is relatively larger.

\section{Effects of the estimation of neutrino energy distribution}

For the calculation of neutrino reaction rates, the energy distribution of (streaming) neutrinos needs to be assumed in our energy-integrated radiation transfer scheme. 
Here, we illustrate the quantitative dependence of the explosion energy on the assumption.

In this work, we assume the Fermi-Dirac-type energy distribution of neutrinos in 
the form of 
\begin{align}
f_{\nu}(\omega) = \frac{1}{e^{\omega/T_\nu-\eta_\nu }+1}, 
\end{align}
where $T_\nu$ and $\eta_\nu$ are parameters to be determined.
For determining them, we use the expression of the energy density of the streaming neutrinos in the comoving frame of the matter
\begin{align}
J = \int \frac{d^3k}{(2\pi \hbar)^3}\ \omega f_\nu(\omega) = \frac{T_\nu^4}{2\pi^2 (\hbar c)^3} F_3(\eta_\nu), \label{eq:estimation1}
\end{align}
where $F_i(\eta)$ is the relativistic Fermi integral of order $i$.
We further assume $T_\nu=T$, i.e., the ``temperature" of streaming neutrinos is assumed to be equal to the local matter temperature.
In our simulation, the absorption and pair-annihilation of neutrinos are calculated using the energy distribution estimated above.
Because the temperature of neutrinos in reality is comparable to the matter temperature in their emission region, which is usually higher than that in their free-streaming region, the assumption of $T_\nu=T$ is likely to introduce an underestimation for the neutrino heating rate to matter (i.e., in the present work, the neutrino heating is conservatively taken into account).

To quantitatively understand the magnitude of the underestimation, we perform a simulation with a different method of the estimation of the energy distribution as follows:
Using the neutrino energy and number luminosity $L_\nu$ and $L_{N,\nu}$, we estimate the neutrino temperature as
\begin{align}
\frac{F_3(0)}{F_2(0)}T_\nu = \frac{L_\nu}{L_{N,\nu}},
\end{align}
where we assumed $\eta_\nu=0$, and defined
\begin{align}
L_\nu &= \int d^3x \sqrt{-g} u_t Q_\mathrm{(leak)},\\
L_{N,\nu} &= \int d^3x \sqrt{-g} u_t \mathcal{R}_{\mathrm{(leak)}}.
\end{align}
Here, $Q_\mathrm{(leak)}$ and $\mathcal{R}_\mathrm{(leak)}$ are the energy and number emissivities in the rest frame of the matter, respectively \citep[for the detail of their definition, see][]{sekiguchi2010a}.

The simulation is performed using the same setup as M032-S075 (and the model is 
referred to as M032-S075-modE).
In this model, the explosion occurs slightly earlier than in M32-S075, reflecting higher heating efficiency due to higher estimated neutrino average energy.
Moreover, the explosion energy in this model is by $\approx 27$\% higher than that for model M32-S075 (see Table~\ref{tab:model}).
This indicates that with our fiducial energy-integrated method the explosion energy may be underestimated by 30\%.

In reality, the systematic error may be even larger due to the 
following reason: 
The neutrino energy distribution estimated in both methods of this paper 
does not depend on the direction.
However, the neutrino temperature should be larger for those emitted from the PNS 
than those from the torus, reflecting the difference of the matter temperature of the 
neutrino sphere, and this causes the angular dependence of the neutrino energy spectrum. 
To take into account such angular dependence of the energy distribution of neutrinos in the energy-integrated scheme, a more elaborated method \citep[e.g.,][]{Foucart2016b} is needed.

\bibliography{reference}
\end{document}